# Dynamic and Continuous Control of Second-Harmonic Chirality through Lithium Niobate Nonlocal Metasurface

Yiwen Liu[1], Chao Meng,[1,*] Sergey I. Bozhevolnyi[1], Fei Ding,[1,2,*]

[1]Centre for Nano Optics, University of Southern Denmark, Campusvej 55, Odense DK-5230, Denmark

[2]School of Electronic Science and Technology, Eastern Institute of Technology, Ningbo, China

*Corresponding author emails: chao@mci.sdu.dk (C.M.); feid@mci.sdu.dk (F.D.)

**Abstract**

Nonlinear chiral light sources are crucial for emerging applications in chiroptics, including ultrafast spin dynamics and quantum state manipulation. However, achieving precise and dynamic control over nonlinear optical chirality with natural materials and metasurfaces, particularly those based on non-centrosymmetric materials such as lithium niobate (LN), is hindered by the complex tensorial nature of the second-order nonlinear susceptibility $\chi^{(2)}$. Here, we demonstrate a nonlinear nonlocal metasurface, comprising plasmonic nanoantennas atop an *x*-cut LN thin film, that enables dynamic and continuous control of second-harmonic (SH) chirality. By leveraging two spectrally detuned resonances arising from the excitation of orthogonally propagating guided modes—enabled by lattice anisotropy and LN birefringence—we achieve full-range tuning of SH chirality, from right- to left-handed circular polarization, simply by rotating the polarization of a linearly polarized pump. The SH chirality, quantified by the Stokes parameter $S_3$, is thereby continuously tuned from 0.991 to −0.993 in simulations and from 0.920 to −0.815 in experiments, while maintaining consistently enhanced SH intensity across the entire tuning range. Our approach opens new avenues for developing compact and tunable chiral sources, with potential applications in integrated nonlinear photonics and adaptable quantum technologies.

**Introduction**

Nonlinear chiral light sources—particularly those capable of emitting circularly polarized (CP) second-harmonic (SH) signals—are essential for emerging applications in ultrafast spin control[1], chiral quantum optics[2], and chiral sensing[3]. The ability to at-source tailor the chirality of second harmonic generation (SHG) would unlock new opportunities for compact, on-chip photonic systems. Despite its importance, the realization of chiral SHG being technically demanding remains poorly explored. Natural nonlinear materials with large intrinsic second-order susceptibilities $\chi^{(2)}$, such as lithium niobate (LN)[4,5], suffer from inherently weak optical nonlinearities, with efficient conversion requiring fulfillment of strict phase-matching conditions that restricts design flexibility[6–8]. Additionally, the complex tensorial nature of $\chi^{(2)}$ hinders the implementation of deterministic control over the polarization state of emitted SH light[9–12]. Consequently, conventional bulk nonlinear sources often require external polarization optics to tailor SH chirality, limiting their compactness and integration potential. These limitations have spurred a growing interest in artificial nonlinear platforms, such as metasurfaces, offering enhanced chiral responses and freedom to design.

Nonlinear metasurfaces (NLMSs) have recently emerged as a powerful platform for enhancing chiroptical effects and generating chiral nonlinear light at subwavelength scales[13–17]. Thus, plasmonic NLMSs incorporating rotational symmetry have enabled chiral SHG under CP excitation by exploiting localized field enhancements and symmetry-driven selection rules in nonlinear optics[18–21]. However, these approaches inherently suffer from low overall efficiencies due to weak nonlinearities, surface-confined subwavelength-sized mode volumes, and high absorption (due to Ohmic losses). Alternatively, all-dielectric metasurfaces, which support low-loss, highly chiral resonant modes, offer a promising route to chiral SHG without the efficiency being limited by material losses[22–24]. In particular, chiral bound states in the continuum (BICs), realized through structural engineering and multipolar interference, offer simultaneous enhancement of SH efficiency and spectrally selective SH chirality by leveraging symmetry-protected high-$Q$ modes with tailored chiroptical response[25–27]. Despite these advances, a critical limitation persists: the chirality of SH emission in most NLMS is fixed upon fabrication, offering no pathways for post-fabrication reconfiguration. To date, there exists no efficient on-chip strategy to achieve active and continuous tuning of SH chirality across the full range from 1 to −1, a capability that is vital for reconfigurable chiral photonics.

In this work, we propose and experimentally demonstrate a nonlocal NLMS, composed of plasmonic nanoantenna arrays integrated atop an *x*-cut LN thin film, that enables dynamic and continuous control of SH chirality directly at the source. By leveraging two orthogonally propagating, spectrally detuned guided-mode resonances (GMRs) that arise from the interplay between LN birefringence and metasurface lattice anisotropy, we achieve near-full-range tuning of the SH chirality, quantified by the Stokes parameter $S_3$, from +0.920 to −0.815 in experiments (+0.991 to −0.993 in simulations), spanning right-handed circular polarization (RCP), through linear polarization (LP), to left-handed circular polarization (LCP). This polarization control is realized solely by rotating the polarization of a linearly polarized pump beam, without altering the device geometry or adding external optical components. Moreover, the meticulously engineered nonlocal GMRs ensure both enhanced and balanced SHG intensity across the entire polarization tuning range, establishing a compact and scalable strategy for tunable chiral light generation in next-generation photonic and quantum technologies.

## Results

### *Design Concept*

The considered nonlocal NLMS consists of a plasmonic rectangular nanoantenna array atop a thin-film *x*-cut LN substrate that supports two spectrally detuned GMRs by virtue of different array periods in orthogonal directions. These nonlocal GMRs tailor the optical anisotropy at the fundamental frequency, enabling efficient and dynamic control of the SH chirality by simply rotating the polarization of an incident (characterized by angle *θ*), linearly polarized fundamental wave (FW) (Fig. 1a). LN is selected as the nonlinear medium not only for its relatively large $\chi^{(2)}$, but also for its pronounced tensor anisotropy, which is often overlooked by exploiting only the dominant $d_{33}$ component. Here, we intentionally harness this anisotropy, in synergy with the engineered GMRs, to realize dynamic and broadband tunability of SH chirality. To dynamically control the SH chirality, the phase difference between two orthogonal in-plane SH polarizations, $\boldsymbol{P}_z^{2\omega}$ and $\boldsymbol{P}_y^{2\omega}$, must be efficiently tuned. According to the intrinsic $\chi^{(2)}$ tensor of LN (see Eq. (S19)), both SH polarizations contain contributions from all three FW electric field components generated inside the LN substrate. Under normal incidence, however, the out-of-plane component $\boldsymbol{E}_x^{\omega}$ is negligible, leaving the SHG primarily governed by $\boldsymbol{E}_z^{\omega}$ and $\boldsymbol{E}_y^{\omega}$. To modulate their relative phase while simultaneously enhancing SH emission, we employ two spectrally detuned GMRs (associated with the array excitation of orthogonally propagating guided modes) as an efficient and controllable modulation mechanism (Fig. 1b-d).

Optimized for a FW near *λ* = 830 nm, the nonlocal NLMS consists of a rectangular lattice of identical gold nanodisks (height H = 50 nm, radius R = 116 nm) patterned on a 200-nm-thick *x*-cut LN slab, with lattice constants $\Lambda_z$ = 436.5 nm and $\Lambda_y$ = 463.5 nm. The 2D array serves to support two GMRs within the LN layer, whereas the nanodisks' localized surface-plasmon resonances lie far outside the spectral range of interest (Fig. S1) and thus have negligible influence. As shown in Fig. 1b, the in-plane birefringence of the *x*-cut LN produces distinct dispersion relations for TE waveguide modes propagating along the *y*- and *z*-directions, denoted $TE_{0,y}$ and $TE_{0,z}$, respectively (see details Supplementary Information S2). To excite these GMRs, the in-plane wave vector of the incident wave, folded by the reciprocal lattice vector of the metasurface, must match the propagation constant of the corresponding guided modes, satisfying parallel momentum conservation[28–30]:

$$k^{\|}_{m_z,m_y} = k^{\|}_{inc} + k^{y}_{m_y} + k^{z}_{m_z} \qquad (1)$$

where $k^{\|}_{inc}$ is the in-plane wave vector of the incident light, $k^{y}_{m_y}$ and $k^{z}_{m_z}$ are the diffracted wave vectors along $y$- and $z$-directions, respectively, and $m_y$ and $m_z$ denote the corresponding diffraction orders. Under normal incidence ($k^{\|}_{inc} = \mathbf{0}$), two orthogonally excited GMRs (GMR$_y$ and GMR$_z$) are formed by adjusting the lattice constants:

$$k^{y}_{m_y} = m_y b_y = \beta_{0,y} \qquad (2)$$

$$k^{z}_{m_z} = m_z b_z = \beta_{0,z} \qquad (3)$$

Here, $\boldsymbol{b}_y = 2\pi/\Lambda_y \cdot \hat{\boldsymbol{y}}$ and $\boldsymbol{b}_z = 2\pi/\Lambda_z \cdot \hat{\boldsymbol{z}}$ are the reciprocal lattice vectors, $m_z$ and $m_y$ are integers, and $\beta_{0,y}$ and $\beta_{0,z}$ denote the propagation constants of the TE$_{0,y}$ and TE$_{0,z}$ fundamental modes, respectively. Leveraging the LN birefringence and lattice anisotropy, GMR$_y$ and GMR$_z$ are deliberately detuned to occur at slightly different wavelengths $\lambda_y$ and $\lambda_z$, set by the respective (grating periodicity, refractive index) pairs $(\Lambda_y, n_z)$ and $(\Lambda_z, n_y)$ (Fig. 1c).

Within the spectral vicinity of these two resonances, rotating the FW polarization by an angle $\theta$ continuously tunes the relative coupling strengths to GMR$_y$ and GMR$_z$ ($\alpha$ and $\beta$ in Fig. 1d). Because GMR$_y$ and GMR$_z$ exhibit dominant in-plane electric field components $\boldsymbol{E}^{\omega}_z$ and $\boldsymbol{E}^{\omega}_y$, respectively, varying $\theta$ modulates their amplitude ratio in a controllable manner. Moreover, as both modes are effectively confined in the LN layer, the enhanced FW local fields ensure efficient SHG. In addition, the intentional spectral detuning between the two GMRs introduces a wavelength-dependent phase difference between $\boldsymbol{E}^{\omega}_z$ and $\boldsymbol{E}^{\omega}_y$. This wavelength dependence creates a design parameter space in which the operating wavelength can be selected to achieve full-range control of SH chirality. Together, these features make the nonlocal NLMS capable of generating high-efficiency SH with near-full-range, in situ control of the SH chirality, allowing for continuous and smooth tuning from RCP through LP to LCP simply by rotating the FW polarization, or equivalently by rotating the device. As a final remark, it is worth noting that the spatially intricate near-field distribution $\boldsymbol{E}^{\omega}_z$ and $\boldsymbol{E}^{\omega}_y$, coupled through the full $\chi^{(2)}$ tensor into $\boldsymbol{P}^{2\omega}_{z,y}$, render the mapping between FW polarization and $S_3$ intrinsically nontrivial. Therefore, achieving access to both chirality extrema ($S_3 = +1$ for RCP and $S_3 = -1$ for LCP) requires a joint optimization of the operating wavelength $\lambda$, FW polarization rotation angle $\theta$, lattice constants ($\Lambda_z$, $\Lambda_y$), and modal detuning.

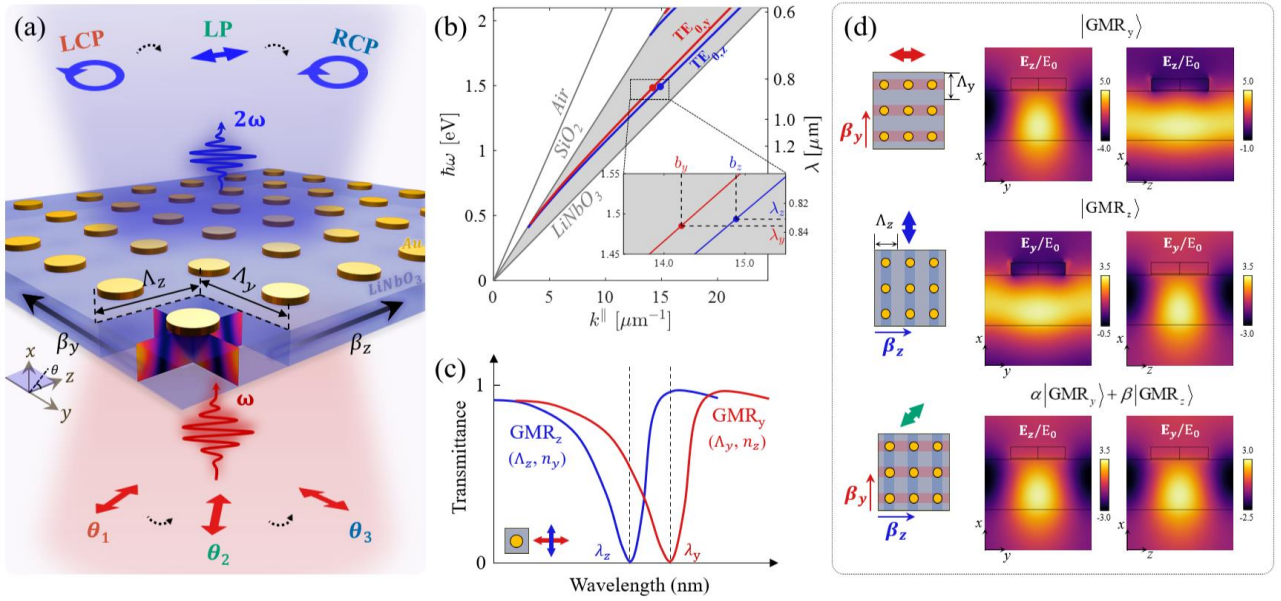

**Fig 1. Design principle of the nonlocal NLMS for dynamic and continuous control of SH chirality.** (a) Schematic of the nonlocal NLMS composed of a plasmonic nanoantenna array atop an *x*-cut LN thin film (optical axis along *z*). Two GMRs originating from orthogonally propagating guided modes enable continuous tuning of SH chirality—from RCP, through LP, to LCP—simply by rotating the polarization (characterized by angle *θ*) of the linearly polarized fundamental wave (FW). (b) Calculated dispersion relations of $TE_{0,y}$ and $TE_{0,z}$ guided modes supported by the anisotropic LN slab waveguide, propagating along the *y*- and *z*-axes, respectively. The intersection points of these modes with the metasurface's reciprocal lattice vectors define two distinct GMR wavelengths $\lambda_y$ and $\lambda_z$. (c) Schematic transmittance spectra of the two orthogonally excited, spectrally detuned $GMR_y$ and $GMR_z$. (d) Polarization-dependent excitation of the two GMRs at the fundamental frequency. At *λ* = 832 nm (near both resonances), simulated electric field distributions show dominant $E_z$ for $GMR_y$ and $E_y$ for $GMR_z$, enabling vectorial FW field control for chirality modulation of the SH emission.

*Amplitude and Phase Tuning of Orthogonal FWs:*

The two principal features of the dual GMRs in our nonlocal NLMS, namely tunable excitation strengths (i.e., *α*, *β*) and spectral detuning, play distinct yet complementary roles in modulating the amplitude and phase differences between the in-plane electric field components $\boldsymbol{E}_z^\omega$ and $\boldsymbol{E}_y^\omega$ at the FW. Figure 2a shows the simulated transmission spectra of the nonlocal NLMS under FW illumination, where the incident electric field is linearly polarized at an angle *θ*, expressed as $\boldsymbol{E}_{in}^\omega = [\boldsymbol{E}_{in,z}^\omega, \boldsymbol{E}_{in,y}^\omega] = E_0[\cos\theta, \sin\theta]$. The transmitted field is decomposed into *z*- and *y*-polarized components $\boldsymbol{E}_{z(y)}^\omega$, with transmittance defined as $T_{z(y)}^\omega = |\boldsymbol{E}_{z(y)}^\omega|^2 / E_0^2$. The FW polarization rotation angle *θ* serves as an effective tuning knob for the transmittance $T_z^\omega$ and $T_y^\omega$ and

thus the relative coupling into GMR$_y$ and GMR$_z$. At $\theta = 0°$ ($\boldsymbol{E}_{in}^{\omega} \parallel z$) and $\theta = 90°$ ($\boldsymbol{E}_{in}^{\omega} \parallel y$), the NLMS transmits only the $z$- and $y$-polarized fields, respectively, at the respective GMR wavelengths (first and third panels in Fig. 2a), leading to exclusive excitation of GMR$_y$ and GMR$_z$. While at intermediate polarization rotation angles (e.g., $\theta = 45°$ or $135°$), both $T_z^{\omega}$ and $T_y^{\omega}$ are nonzero, indicating simultaneous excitation of the two GMRs with tunable relative couling strengths (second and forth panels in Fig. 2a). Moreover, the two GMRs are spectrally detuned, with $\lambda_y = 833.5$ nm GMR$_y$ and $\lambda_z = 830.0$ nm GMR$_z$, and exhibit markedly different quality factors: $Q_z \approx 244$ for GMR$_z$ versus $Q_y \approx 54$ for GMR$_y$. This $Q$-factor asymmetry originates from their different lattice constants, which control the spatial density of the gold nanodisks and thereby the coupling strength between the plasmonic metasurface and the guided modes. A reduced lattice period increases mode confinement in the LN waveguide, yielding a higher-$Q$ resonance. The intentionally higher-$Q$ GMR$_z$ enhances the $\boldsymbol{E}_y^{\omega}$ component, compensating for the intrinsically weaker $\chi^{(2}$ tensor elements along the $y$-direction in LN.

Importantly, both GMRs exhibit near-zero transmittance at their respective resonant wavelengths, ensuring that rotation of the FW polarization rotation angle $\theta$ effectively redistributes optical power between the $\boldsymbol{E}_z^{\omega}$ and $\boldsymbol{E}_y^{\omega}$ components within the spectral window between them. The normalized amplitude difference $\Delta A$, which quantifies the relative strength of these orthogonal field components, is defined as:

$$\Delta A = \frac{|\boldsymbol{E}_z^{\omega}| - |\boldsymbol{E}_y^{\omega}|}{\sqrt{|\boldsymbol{E}_z^{\omega}|^2 + |\boldsymbol{E}_y^{\omega}|^2}} \tag{4}$$

where $\boldsymbol{E}_z^{\omega}$ and $\boldsymbol{E}_y^{\omega}$ represent the far-field components, numerically computed using COMSOL (see Method: Numerical Simulations for details). As shown in Fig. 2b-c, $\Delta A$ varies continuously from $-1$ to $+1$ across both [0°, 90°] and [90°, 180°] intervals of $\theta$ for any wavelength in the spectral range from 825 to 840 nm, exhibiting mirror symmetry about $\theta = 90°$.

In addition to controlling the amplitude difference, introducing a tunable phase difference $\Delta\varphi = \varphi_z - \varphi_y$ is equally essential. As shown in Fig. 2d, $\Delta\varphi$ varies by approximately 150° across the examined wavelength range. A discrete 180° phase jump appears when $\theta > 90°$, originating from the sign reversal of the incident-field component $\boldsymbol{E}_{in,z}^{\omega}$. This discontinuity leads to distinct $\Delta\varphi$ behaviors in the two intervals [0°, 90°] and [90°, 180°], even though $\Delta A$ spans the full range $[-1, 1]$ in both cases. As a result, the accessible SH chirality in these two intervals is not mirror-symmetric, enabling a broader tuning range. Figure 2e further details the spectral phase profiles of GMR$_z$ and GMR$_y$ at $\theta = 45°$, where their spectral detuning induces significant phase differences ranging from approximately −50° to 100°.

Although the anisotropic nonlocal NLMS design with dual detuned GMRs enables independent tuning of both amplitude and phase differences between orthogonal far-field FW components, a specific amplitude-phase combination at the FW does not directly translate into a simple one-to-one map of the resulting SH chirality.

This is because the nonlinear polarization is generated locally within the LN layer, and at resonance, the FW near fields exhibit complex spatial distributions (see details in Supplementary Information S3). As a result, the local nonlinear polarization $P_z^{2\omega}(r)$ varies across the unit cell, and the SH far-field emission is the coherent superposition of these spatially distributed sources (Fig. S3). Thus, while far-field FW simulations offer valuable insights into the integrated signature of the underlying fields, they cannot capture the full pointwise interplay of $E_z^\omega(r)$ and $E_y^\omega(r)$, nor their respective contributions to the nonlinear response. These interactions must be resolved with numerical nonlinear simulations. In the following section, we show that, despite this complexity, the tunable far-field FW behavior is faithfully inherited by the dynamically modulated SH chirality, validating it as a predictive metric for nonlinear polarization control.

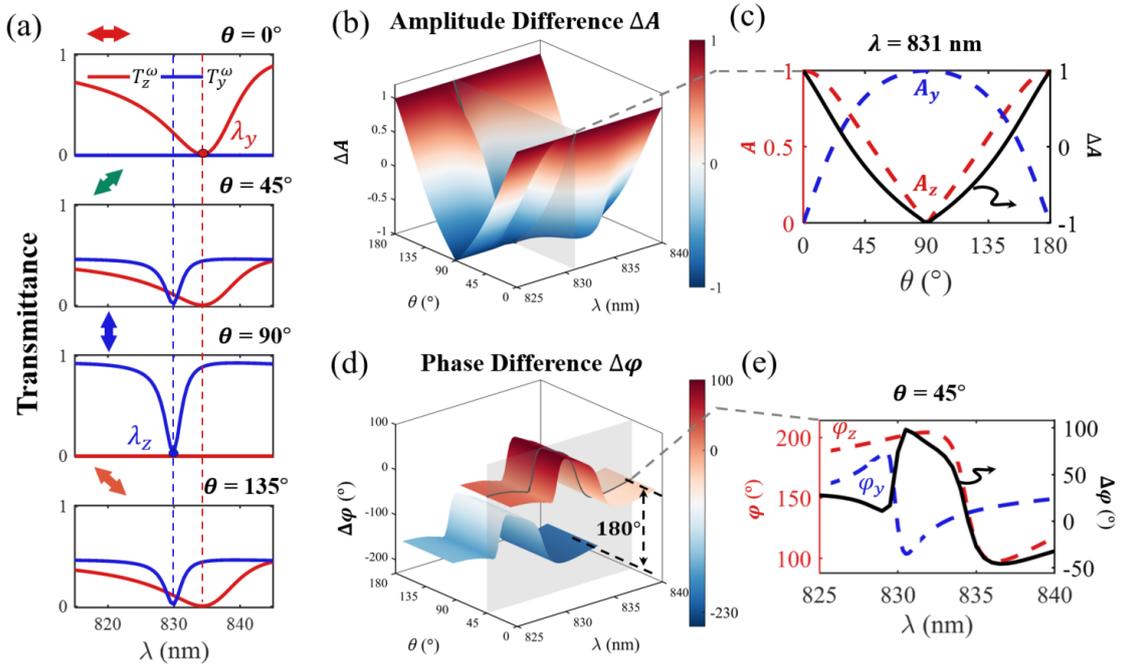

**Fig 2. Simulated complex transmission coefficients of the nonlocal NLMS at FW**. (a) Simulated transmission spectra under varying FW polarization rotation angles $\theta$. The red and blue curves correspond to the transmittance projected onto the $z$- and $y$-polarization channels, denoted as $T_z$ and $T_y$, respectively. (b) Simulated amplitude difference $\Delta A$ as a function of the FW wavelength $\lambda$ and polarization rotation angle $\theta$. (c) Normalized amplitude $A_z = |E_z^\omega|/(|E_z^\omega|^2 + |E_y^\omega|^2)^{1/2}$, $A_y = |E_y^\omega|/(|E_z^\omega|^2 + |E_y^\omega|^2)^{1/2}$, and amplitude difference $\Delta A$ as a function of $\theta$ at a fixed FW wavelength $\lambda = 831$ nm. (d) Simulated phase difference $\Delta\varphi$ as a function of FW wavelength $\lambda$ and polarization rotation angle $\theta$. A 180º phase discontinuity emerges at $\theta = 90$º, consistent with the expected π-phase shift associated with linear polarization inversion. (e) Simulated spectral phase responses of the GMR$_z$ and GMR$_y$ (red and blue lines) and their phase difference (black line) at $\theta = 45$º.

*Dynamic and Continuous Control of SH Chirality: Simulation*

Motivated by the far-field FW behavior, one might intuitively expect that achieving full-range modulation of the amplitude difference together with broad tunability of the phase difference between two orthogonal FW components would naturally produce a wide range of SH chirality. However, as discussed above, the SH far-field chirality cannot be directly inferred from FW properties alone and necessitates nonlinear simulations for accurate prediction. To realize continuously tunable SH chirality covering both RCP and LCP states, we optimized the metasurface parameters by directly targeting the desired SH performance. As an initial step, we employed an effective nonlinear susceptibility tensor model[32–34] (see details in Supplementary Information S4), which significantly reduces computational cost by avoiding extensive scans over the FW polarization rotation angle $\theta$. This design was then refined with nonlinear simulations under the undepleted pump approximation to leverage the detailed near-field interactions (see Method: Numerical Simulations for details). Notably, our nonlocal NLMS design also explicitly accounts for the strong $\chi^{(2)}$ anisotropy of LN, which differently enhances SH emission in orthogonal polarization channels, thereby maintaining stable SH output intensity throughout the chirality tuning process.

Among all design parameters, the lattice periodicities predominantly determine the spectral separation between the two GMRs, while the nanodisk radius provides fine spectral tuning. With the optimized geometry, both extreme chiral states ($S_3 = \pm 1$) can be reached by varying $\theta$. Fig. 3a shows the simulated $S_3$ map as a function of ($\lambda$, $\theta$), revealing smooth, wide chirality modulation governed by the two detuned GMRs in the nonlocal NLMS. In contrast, a bare 200-nm-thick $x$-cut LN film exhibits negligible chirality tunability, with $S_3$ constrained to [−0.2, 0.2] for all $\theta$ (Fig. S5 and S6), owing to the minimal phase difference between $\boldsymbol{E}_z^\omega$ and $\boldsymbol{E}_y^\omega$ and hence between $\boldsymbol{P}_z^{2\omega}$ and $\boldsymbol{P}_y^{2\omega}$. At the optimal FW wavelength $\lambda = 831$ nm, our nonlocal NLMS enables near-full-range, continuous control of SH chirality solely by rotating the linear pump. The simulated $S_3$ (solid line in Fig. 3b) spans from $S_3^{max} = +0.991$ at $\theta = 58º$ to $S_3^{min} = -0.993$ at $\theta = 133º$, corresponding to an absolute modulation depth of $\Delta S_3 = 1.984$ (Fig. 3b). Multiple intermediate LP states ($S_3 = 0$) also appear, with one representative point marked in Fig. 3b. Notably, the $S_3$ ($\theta$) curve predicted by the effective tensor theory (dashed line in Fig. 3b) closely matches the nonlinear simulation, validating the model's predictive capability. In this approach, the subwavelength unit cell is homogenized to have an effective transverse $\chi^{(2)eff}$ that induces a uniform nonlinear polarization density, thereby linking the FW far-field response to the resulting SH chirality (see details in Supplementary Information S4). The corresponding evolution of SH polarization states is visualized via full Stokes parameter mapping on the Poincaré sphere (Fig. 3c), demonstrating deterministic and reversible control over SH chirality, which stands in clear contrast to the weak tunability of the bare LN film ( Fig. S6). Furthermore, the NLMS exhibits spectral robustness in the vicinity of the working wavelength, where the $S_3(\theta)$ curves maintain similar profiles within the FW range from 831 to 834 nm despite a modest decrease in peak values (Fig. S7a). This robustness is essential for experimental implementation, where the pump typically has a finite linewidth rather than being perfectly monochromatic.

In addition to chirality tunability, maintaining both enhanced and balanced SH intensity throughout the tuning range is essential for practical applications. However, the large disparity among $\chi^{(2)}$ elements in bare LN films—where $|d_{33}|$ is much larger than $|d_{31}|$, $|d_{22}|$, and $|d_{21}|$, biases the SH signal strongly toward the $z$-direction, leading to pronounced SH intensity variation with varied FW polarization rotation angles. In our NLMS, we mitigate this imbalance through two complementary strategies: (i) increasing the $Q$-factor of GMR$_z$ relative to GMR$_y$ to boost the local FW electric field $E_y$; and (ii) slightly shifting the operating wavelength towards $\lambda_z$ = 830 nm, where GMR$_z$ dominates. These optimizations ensure substantial near-field overlap of the two orthogonal GMRs across the tuning range, producing comparable $P_z^{2\omega}$ and $P_y^{2\omega}$ components and thus delivering an enhanced, balanced SH signal. As shown in Fig. 3d, the total SH intensity is significantly higher than that of the bare LN film over the tuning range, with a maximum enhancement factor of over 300. More importantly, at the optimal FW wavelength of 831 nm, the SH intensity remains stable as the FW polarization rotation angle $\theta$ varies (Fig. 3e and 3f), showing moderate changes even as the chirality switches from RCP to LCP (red and blue star in Fig. 3f). The ability to sustain enhanced, balanced SH intensity throughout chirality tuning underscores the advantages of our nonlocal NLMS for chiral photonics and nonlinear optical systems.

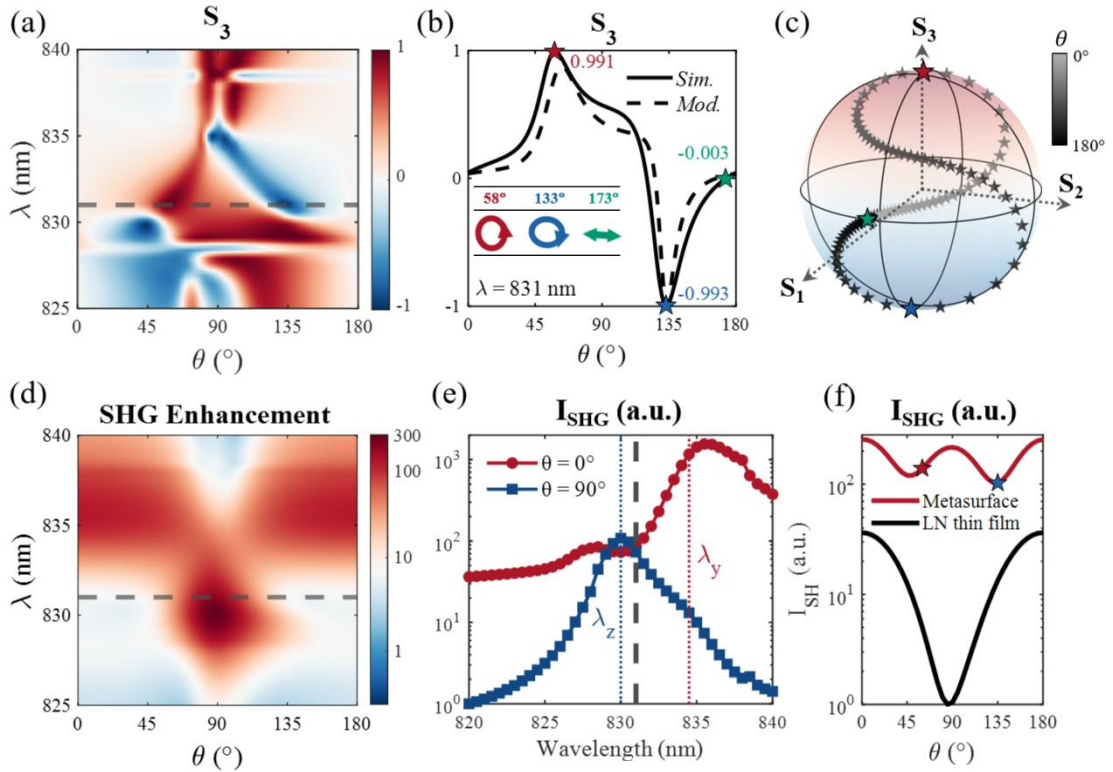

**Fig 3. Numerical analysis of dynamic and continuous control of SH chirality.** (a) Calculated SH chirality $S_3$ mapped as a function of FW wavelength $\lambda$ and polarization rotation angle $\theta$. The grey dashed line at $\lambda$ = 831 nm indicates the wavelength corresponding to (b) and (c). (b) Calculated SH chirality $S_3$ as a function of incident FW polarization rotation angle $\theta$ at $\lambda$ = 831 nm. The solid and dashed lines indicate results from numerical simulation and the effective nonlinear susceptibility tensor model, respectively. Representative RCP, LP, and LCP polarization states are explicitly marked with red, green and blue stars. (c) Evolution of SH polarization on the Poincaré sphere as $\theta$ increases from 0º (light color) to 180º (dark color). (d) Calculated SHG enhancement of the nonlocal NLMS relative to a bare *x*-cut LN film, mapped as a function of FW wavelength $\lambda$ and polarization rotation angle $\theta$. The grey dashed line indicates $\lambda$ = 831 nm. (e) Calculated SH intensity as a function of FW wavelength $\lambda$ at $\theta$ = 0º and 90º, normalized to the minimum value among all data points. The grey dashed line indicates $\lambda$ = 831 nm. (f) Calculated SH intensity at $\lambda$ = 831 nm as a function of $\theta$, normalized to that of the bare LN film at $\theta$ = 90º.

*Dynamic and Continuous Control of SH Chirality: Experiment*

To experimentally validate the numerically predicted tunability of SH chirality, we fabricated the optimized nonlocal NLMS and performed both linear and nonlinear optical characterizations. The NLMSs were fabricated using electron beam lithography (EBL), followed by metal deposition and lift-off (see details in Method: Fabrication). Insets of Fig. 4a present an optical microscope image of the fabricated 60 μm × 60 μm array and a zoomed-in scanning electron microscope (SEM) images of the gold nanodisks, with the nanodisk dimensions and lattice periodicities closely matching the design. Additional SEM images are provided in Fig. S7. Following the fabrication, we built a microscopy platform for wavelength-resolved, full-Stokes linear and nonlinear optical characterizations (Fig. 4a, and see details in Method: Characterization and Fig. S8), where a mode-locked femtosecond titanium-sapphire laser (Tsunami 3941-X1BB, Spectra-Physics) served as the excitation source, with its measured spectrum shown in Fig. 4b (black solid curve). Figure 4c compares the designed and measured linear transmission spectra under *z*- and *y*-polarized FW excitations, showing good agreement in both GMR wavelengths and overall spectral profiles. The measured resonances occur at $\lambda_z$ = 829 nm and $\lambda_y$ = 833nm, close to the design values of 830 and 833.5 nm, respectively. Such small deviations have a minimal impact on the resonance spacing, which is crucial for SH chirality control. The measured full width at half maximum (FWHM) and *Q* factors are 7.3 nm ($Q \approx 113$) for GMR$_z$ and 17.1 nm ($Q \approx 49$) for GMR$_y$, respectively, compared with simulation values of 3.4 nm ($Q \approx 244$) and 15.3 nm ($Q \approx 54$), respectively. The reduced *Q*-factors are attributed to increased material absorption and surface roughness in the fabricated gold nanostructures, as well as the finite size of the fabricated nonlocal NLMS array[35].

To probe the SH chirality at different wavelengths while suppressing other $\chi^{(2)}$ processes, such as sum-frequency generation, the femtosecond pump (FWHM = 17.64 nm) was spectrally narrowed using a laser-line band-pass filter (BPF, FWHM ≈ 2.8 nm, black dotted line in Fig. 4b) and tuned in the central wavelength by tilting the BPF. Although the filtered FW pump is not strictly monochromatic, its narrowed bandwidth is sufficient to resolve the predicted chirality-tuning behavior. At FW $\lambda$ = 828 nm, the measured SH $S_3(\theta)$ curve exhibits near-full chirality tuning, ranging from $S_3^{max}$ = 0.920 at $\theta$ = 60° to $S_3^{min}$ = −0.815 at $\theta$ = 120°, as shown in Fig. 4d. To our knowledge, this constitutes the first experimental demonstration of reversible and continuous switching of SH chirality, achieving a record-high modulation depth of $\Delta S_3$ = 1.735. The measured $S_3(\theta)$ curve shows minor deviations from the simulations in shape and in the positions of extrema, likely due to differences in the GMR lineshapes. Particularly, the transmittance minima in the experiment does not reach zero, slightly altering the $\Delta A(\theta)$ mapping. Nevertheless, the experimentally retrieved polarization trajectory on the Poincaré sphere (Fig. 4e) traces a characteristic figure-eight pattern predicted in simulations (Fig. 3c), confirming continuous tuning between RCP, LP, and LCP states. The trajectory initiates near the equator, moving upward counterclockwise, and then downward clockwise. Representative SH spectra measured in the RCP and LCP bases at the points of maximal chirality are shown in Fig. 4f, revealing high-contrast chiral emission. Full Stokes analysis yields an average degree of polarization (DoP, defined as DoP = $\sqrt{S_1^2 + S_2^2 + S_3^2}$) of 0.959 across the entire tuning range, indicating high-purity SH polarization (Fig. S10). Additional measurements at neighboring FW wavelengths (Fig. S9) confirm predicted SH chirality tuning. Importantly, the nonlocal NLMS maintains enhanced and stable SH intensity across the entire $\theta$ range (Fig. 4g), owing to the balanced field enhancement engineered between the two detuned GMRs. However, the measured SH intensity is approximately nine times lower than in design simulations (Fig. 3f), primarily due to reduced $Q$-factors of the GMRs, which can be improved by utilizing large-scale fabrication techniques[36] and the use of high-quality, low-loss materials[37].

As a final remark, it is worth noting that our nonlocal NLMS platform can be extended to anisotropic meta-atoms such as brick-shaped elements with additional degrees of freedom (Fig. S12). By adjusting the nanobrick length and width, the radiative losses and hence the linewidths of the two GMRs can be selectively tailored, thereby enabling finer control over their spectral phase profiles and the phase difference between orthogonal FW components.

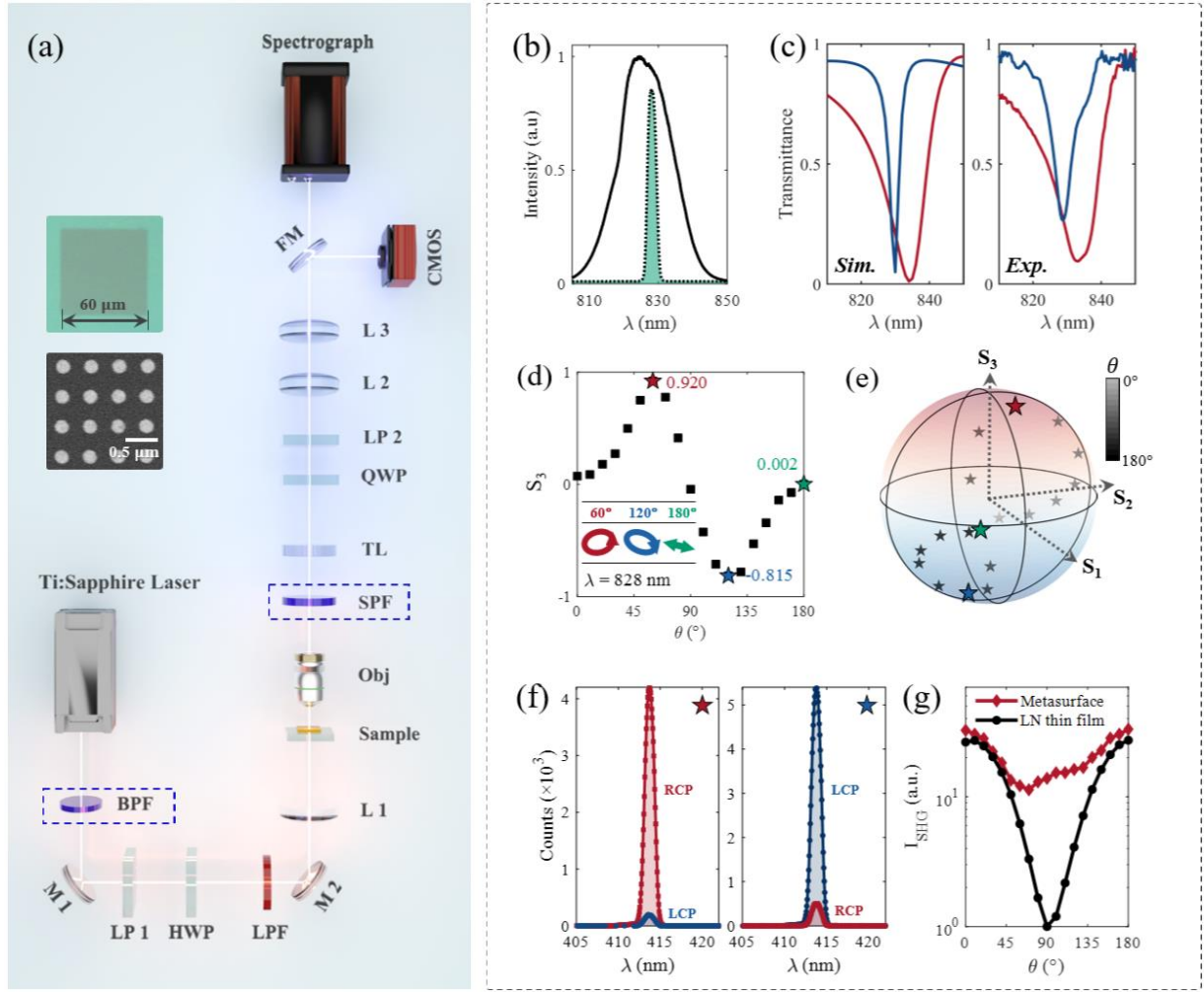

**Fig 4. Experimental demonstration of dynamic and continuous control of SH chirality.** (a) Schematic of the optical setup used for both linear and nonlinear measurements. Additional optical components (blue dashed box) are included specifically for SHG characterization. Insets: Optical microscope and SEM images of the fabricated MS array. (b) Normalized spectra of the femtosecond laser (black solid) and the spectrally filtered laser (black dotted with green fill). The filtered spectrum is centered at 828 nm with FWHM ≈ 2.8 nm. (c) Simulated and experimentally measured linear transmission spectra under pump polarization along the *z*- (red) and *y*- (blue) axes. (d) SH $S_3(\theta)$ curve at FW $\lambda$ = 828 nm, demonstrating continuous tuning of SH chirality with pump polarization rotation. Representative polarization states of RCP, LP, and LCP are illustrated by polarization ellipses. (e) SH polarization trajectory on the Poincaré sphere as $\theta$ increases from 0º to 180º, shown with a light-to-dark color gradient. Polarization states of RCP, LP, LCP are marked along the trajectory. (f) Measured SH spectra projected to the RCP and LCP bases at the points of maximal and minimal $S_3$ values in panel (d), highlighting the strong chirality contrast. (g) SHG intensity at FW $\lambda$ = 828 nm as a function of polarization rotation angle $\theta$ normalized to that of the bare LN film at $\theta$ = 90º.

**Conclusion**

In summary, we have demonstrated a nonlocal NLMS, comprising an *x*-cut LN thin film integrated with a plasmonic rectangular nanoantenna array, that enables dynamic and continuous control of SH chirality. This capability arises from the engineered superposition of two spectrally detuned GMRs, originating from the excitation of two orthogonally propagating guided modes due to the interplay between lattice anisotropy and LN birefringence, that enables tailoring the in-plane FW electric field distribution. Through the vectorial nonlinear response governed by the LN anisotropic $\chi^{(2)}$ tensor, the SH chirality can precisely be tuned simply by rotating the pump polarization. Consequently, the SH chirality can continuously be varied from RCP to LCP, corresponding to simulated $S_3$ values varying from +0.991 to −0.993 and experimental values from +0.920 to −0.815, while preserving high polarization purity and maintaining consistently enhanced and balanced SH intensity across the entire tuning range. We believe that the reported approach to compact, at-source tunable nonlinear chiral light sources opens new opportunities for integrated nonlinear photonics, chiral optics, and adaptable quantum technologies.

**Methods**

**1. Fabrication**

The fabrication of the MS was carried out using electron-beam lithography (EBL), thermal evaporation, and lift-off techniques. The structures were patterned on a commercially sourced *x*-cut lithium niobate ($LiNbO_3$) thin film with a thickness of 200 nm, which was bonded to a 2.5 μm-thick silicon dioxide ($SiO_2$) buffer layer atop a 500 μm-thick quartz substrate (NanoLN, Jinan Jingzheng Electronics Co.). In the EBL process, a 100 nm-thick PMMA layer (2% in anisole, MicroChem) was spin-coated onto the $LiNbO_3$ surface and baked at 180 °C for 2 min, after which a 40 nm-thick conductive-polymer layer (AR-PC 5090, Allresist) was spin-coated and baked at 90 °C for 1 min. Pattern definition was then performed at an acceleration voltage of 30 keV. After exposure, the conductive-polymer layer was removed in deionized water, and the exposed PMMA was developed in a solution containing methyl isobutyl ketone (MIBK) and isopropyl alcohol (IPA) at a ratio of MIBK to IPA of 1:3 for 35 s, followed by immersion in an IPA bath for 60 s. Next, a titanium adhesion layer (2 nm) and a gold layer (50 nm) were sequentially deposited by thermal evaporation. Finally, a lift-off process was performed to remove the remaining resist and excess metal, yielding the desired gold nanostructures.

**2. Numerical Simulations**

All the simulations in this work were conducted with the Electromagnetic Waves, Frequency-Domain (EWFD) module of COMSOL Multiphysics 6.3 based on the finite-element method. (i) Fundamental-wave (FW) simulation. A normally incident plane wave was injected from the silica substrate via a periodic port, with lateral Floquet periodic boundary conditions applied to emulate an infinite array. The complex amplitudes of the transmitted FW components $E_z^\omega$ and $E_y^\omega$ were integrated over the unit cell on a horizontal monitor plane on the air side. (ii) Second-harmonic generation (SHG) simulation—two-step EWFD scheme. The first step is simulated at the fundamental frequency with EWFD to retrieve the local FW field distributions $E^\omega(r)$. In the second step, the resulting nonlinear polarization distribution $P^{2\omega}(r)$ induced inside the LN thin film was evaluated and used as the source term in a separate frequency-domain simulation at the SH frequency, yielding the SH fields. The resulting SH far-field electric components $E_z^{2\omega}$ and $E_y^{2\omega}$ were integrated over the unit cell on a horizontal monitor plane in the air, from which the SH polarization state was determined.

## 3. Characterization

In both FW and SH characterizations, a mode-locked femtosecond titanium-sapphire laser (Tsunami 3941-X1BB, Spectra-Physics; ~ 65 fs, 82 MHz repetition rate) was employed as the excitation source, delivering pulses with tunable central wavelengths ranging from 750 to 900 nm. Prior to entering the optical setup, the beam was coupled into two cascaded 1-meter-long single-mode fibers (P3-780PM-FC-1), ensuring a Gaussian-like output profile and while introducing dispersion that broadened the pulse duration to ~ 3 ps. The linearly polarized laser beam was incident from the substrate side of the sample. As the finite waist of a Gaussian beam influences the excitation of GMRs[38], a beam spot diameter of approximately 40 µm was selected—slightly smaller than the MS array, yet sufficiently large to ensure that the beam's angular divergence remains relatively low, thereby maintaining efficient excitation. Prior to measurements, normal incidence illumination was carefully ensured by precise mechanical alignment, owing to the angular sensitivity of GMRs (see Fig. S13). The beam was focused on the sample using a lens with a focal length of 5 cm. A long-pass filter (cut-off: 550 nm) was placed in front of the sample to remove unwanted shorter-wavelength components. After transmission through the sample, the light was collected using a 50× objective (NA = 0.55). An iris was positioned at the first image plane to filter out the area of interest in the measurement. A flip lens and a flip mirror were used to direct the beam into two separate detection paths, sending the real-plane image to a CCD and the Fourier-plane image to a spectrometer (Andor Kymera 328i). For SH measurements, a laser-line bandpass filter (BPF, FWHM ≈ 2.8 nm) was inserted to narrow the spectral bandwidth of the FW pump, whose central wavelength could be tuned by in-plane titling (see Fig. S14). A short-pass filter (SPF, cut-off: 750 nm) was placed after the sample to eliminate the residual fundamental beam. Additionally, a linear polarizer (LP) in combination with a quarter-wave plate (QWP) was used to project the SH signal onto six polarization bases: horizontal ($I_x$), vertical ($I_y$), 45º linear ($I_a$), 135º linear ($I_b$), right-circular ($I_r$), and left-circular ($I_l$). Based on these measurements, the full Stokes parameters were calculated using the standard definitions: $S_1 = (I_x - I_y) / (I_x + I_y)$, $S_2 = (I_a - I_b) / (I_a + I_b)$, $S_3 = (I_r - I_l) / (I_r + I_l)$.

**Acknowledgments**

This work was supported by the Villum Fonden (Grant No. 37372 and 50343), the Carlsberg Foundation (CF24-1777), and Danmarks Frie Forskningsfond (1134-00010B).